\documentclass[12pt,preprint]{aastex}
\usepackage{epsfig}
\shorttitle{H$\alpha$ equivalent width difference}
\shortauthors{Cedr\'es et al.}
\begin{document}
\title{Comparison of the H$\alpha$ equivalent width of  H~{\sc ii} regions in a flocculent and a grand design galaxy: possible evidences for IMF variations}
\author{Bernab\'e Cedr\'es and Jordi Cepa \altaffilmark{1}} 
\affil{Instituto de Astrof\'{\i}sica de Canarias, E-38200 La Laguna, Tenerife, Spain}
\and 
\author{Akihiko Tomita}
\affil{Faculty of Education, Wakayama University, Wakayama 640-8510, Japan}
\altaffiltext{1}{Departamento de Astrof\'{\i}sica, Facultad de F\'{\i}sica, Universidad de La Laguna, E-38071 La Laguna, Tenerife, Spain}
\email{bce@iac.es}
\begin{abstract}
We present here a study of the H$\alpha$ equivalent widths of the flocculent galaxy NGC~4395 and the grand design galaxy NGC~5457. 
A difference between the mean values of the H$\alpha$ equivalent widths for the two galaxies
 has been found. Several
 hypotheses are presented in order to explain this difference: differences in age, metallicity, star formation rate, photon leakage and 
 initial mass function. Various 
 tests and Monte Carlo models are used to find out the most probable cause of this difference. The results show that the possible
 cause for the difference could be a variation in the initial mass function. This difference is such that it seems to favor a fraction of more 
 massive stars in the grand design galaxy when compared with the flocculent galaxy. This could be due to a change of the environmental conditions due to a density wave.
\end{abstract}

\section{Introduction}
 
The universality or the variation of the IMF is a controversial matter. There are evidences that some star clusters in starbursts galaxies
 have a fraction of massive stars larger than those predicted by a standard IMF (for example: McCrady, Gilbert \& Graham 2003, F\"orster Schreiber et al. 2003,
 Smith \& Gallagher 2001, Alonso-Herrero et al. 2001, Sternberg 1998). Elmegreen (2004) suggests that this increase in the fraction of 
massive stars could be due to a coalescence of dense pre-stellar condensations in denser clusters. Shadmehri (2004) presents a model of 
the IMF that would result from this hypothesis.

Another good experimental sites for the gas density variation are the spiral arms. Lin \& Shu (1964) and Seiden \& Gerola (1982) laid down the theoretical bases about the spiral structure. Since these early works, 
both the Spiral Density Wave (SDW) theory and the Self Stochastic Propagation of Star Formation (SSPSF), respectively, have matured 
quite fast.
 
From broad band imaging of a sample of galaxies, Elmegreen \& Elmegreen (1984) suggested that grand design galaxies have large
 enhancements of stellar density in the spiral arms, while the arms of flocculent galaxies were mainly outlined by star formation.
 Then grand design spirals could be density wave dominated, while flocculent spirals would mainly host propagating star formation 
processes in their discs. In fact SSPSF models can only reproduce arms of flocculent morphology unless a density wave is also present, 
as shown by Gonz\'alez-P\'erez \& Cepa (2005). From a comparison of different star formation rate (SFR) tracers vs. arm type for a sample
 of spirals, Elmegreen \& Elmegreen (1986) did not found any differences in the star formation rates averaged over the whole galaxy discs,
claiming that SDW do not trigger star formation in grand design spirals in comparison with flocculent ones. The concept of triggering is 
intended as an enhancement of SFR over that determined by the gas enhancements due to the DW present in the spiral arms. Similar results 
were found by McCall \& Schmidt (1986) from SN samples. 

On the contrary, other researchers such as Sitnik (1989,1991), Efremov (1985) and Hodge et al.\ (1990), claimed that there was star 
formation triggering evidence in some spirals. Moreover, comparing H$\alpha$ luminosity with CO and H~{\sc i} surface densities, Cepa \&
 Beckman (1989, 1990a,b), Tacconi \& Young (1990) and Knapen et al. (1992) found evidences pointing to a substantially larger star 
formation efficiency (SFE) in the arms with respect to star forming regions in the interarm disc, i.e.: spiral arms are more efficient 
transforming gas into stars. However, with similar spatial resolution, molecular gas density enhancements are larger than broadband UV or
 blue surface brightness enhancements, a result that would support non-triggering.

The possibility of spiral arms favouring the 
formation of a larger fraction of massive stars, i.e.: changing the IMF but keeping the total SFR at similar levels that the interarm 
disc, could harmonize the previous observational results of larger SFE in grand design spirals but with similar overall SFR than 
flocculent spirals. Since spiral arms of grand design spirals are sites of higher density due to the 
passage of the density wave, it might result in a top-heavy IMF in the arms with respect to the rest of the disc, although probably the 
result would not be so noticeable as in the case of starbursts galaxies. Thus, both theoretical results and observations seem to indicate a possible IMF shift 
towards more massive stars in higher density regions.
 
In this paper, we present a comparative study of a grand design and a flocculent galaxy that reveals a difference in the H$\alpha$ 
equivalent widths that, after discarding other possible factors, seems to suggest the presence of possible IMF variations on the grand 
design spiral with a production of a higher fraction of massive stars.

In section 2 we describe the data. In section 3 we explain the results obtained from the 
H$\alpha$ equivalent width and we make a comparision between our two galaxies. In sections 4 and 5 we discuss the effects of the age 
and the abundace on the H$\alpha$ equivalent width distribution, respectively. In section 6 we describe our models and the Monte Carlo 
tests developed to interpret the difference in the equivalent widths. In sections 7 and 8 are discussed the IMF and the photon leakage 
effects, respectively.

\section{The dataset}

The dataset used in this analysis has been presented in a previous paper (Cedr\'es \& Cepa 2002). The gathering, 
reduction, and corrections 
applied to the data are described there. Some of the applied corrections compensated for galactic and extragalactic extinction,
the underlying absorption due massive stars in H~{\sc ii} regions,  contamination by [N~{\sc
ii}]$\lambda\lambda$6548, 6584 lines, and  emission from the underlying galactic disk.  This last correction is critical because it will 
wipe out all the continuum contribution of the disk to the H~{\sc ii} regions, leaving only the intrinsic continuum for each region. 

We have selected a grand design galaxy, NGC~5457, with arm class type 9 according to the Elmegreen \& Elmegreen (1987)
classification. This galaxy was chosen for its proximity,  its large angular size and low inclination, and its richness
in H~{\sc ii} regions. For this
galaxy, the compiled data consist of 338 H~{\sc ii} regions of the inner parts, including the 35\% of the disk: 214 arm 
regions and 124
inter-arm regions. The method employed  to determine whether an H~{\sc ii} region belongs to an arm or  an inter-arm 
zone is described
in Cedr\'es \& Cepa (2002).

We have also selected a flocculent galaxy, NGC~4395, with arm class type 1 (Elmegreen \& Elmegree 1987). This galaxy was
selected for its very intense star-forming processes and its very bright H~{\sc ii} regions in H$\alpha$, [O~{\sc ii}]
and [O~{\sc iii}]. For this galaxy the compiled data consist of 158 H~{\sc ii} regions over the whole disk, and
 we  made no distinction between arm and inter-arm regions because the galaxy lacks a clear arm design.

The parameters of the two galaxies are summarized in Table \ref{gal}.

\begin{table}
\caption{Galaxy parameters}
\begin{tabular}{c|c|c|c|c|c}
Galaxy & Morphological & Distance & Arm & $M_{B}$ & Number of \\
name & type & (Mpc)& class & & regions \\
 & (1) &  & (2) & (3) &  \\
\hline
NGC 5457 & SAB(rs)cd & 7.4 (4)& 9 & 8.31 & 338 \\
NGC 4395 & SA(s)m & 4.6 (5)& 1 & 10.64 & 158 \\
\hline
\end{tabular}
\label{gal}
\\
(1) De Vaucouleurs et al.\ (1991).\\
(2) Elmegreen \& Elmegreen (1987).\\
(3) http://nedwww.ipac.caltech.edu/ \\
(4) Kelson et al.\ (1996).\\
(5) Karachentsev et al.\ (2003).\\
\end{table}

\section{H$\alpha$ equivalent width}

The equivalent width of the H$\alpha$ emission line is defined according to
\begin{equation}
{\rm EW}_{\rm H\alpha}=\frac{F_{\rm H\alpha}}{F_{\rm cH\alpha}}a,
\end{equation}
where $F_{\rm H\alpha}$ is the flux in H$\alpha$ from the H~{\sc ii} region, $F_{\rm cH\alpha}$ is the flux of the 
continuum,
and $a$ is the FWHM of the continuum filter in \AA. In this way, we obtained the ratio of H$\alpha$ emission to the flux
in a 1 \AA\ bandpass of the underlying stellar continuum (Belley \& Roy 1992). This magnitude gives a measure of the ratio 
of ionizing photons
from massive stars to the photons of the continuum from the embedded stars. It is therefore sensitive to variations in 
the IMF, 
the evolutionary stage of the ionizing stars, and the metallicity (Leitherer et al.\ 1999; Kennicutt et al.\ 1989; Martin 
\& Friedli 1999; 
Bresolin \& Kennicutt 1997).

\begin{figure}
\epsfig{file=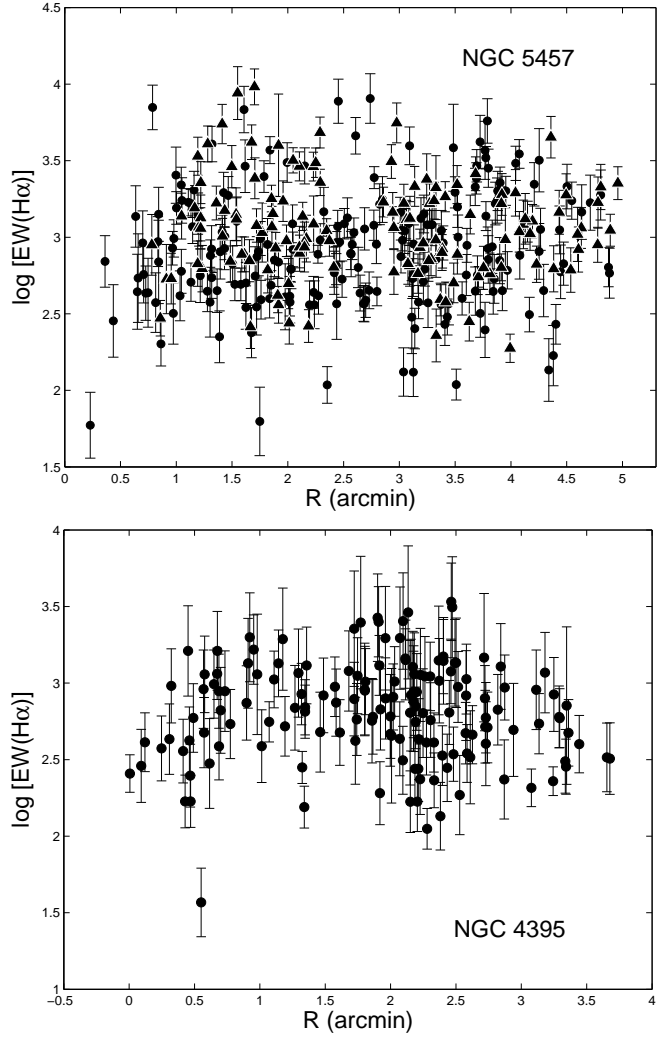,width=88mm}
\caption{H$\alpha$ equivalent width as a function of galactocentric radius in arcminutes for NGC~5457 (upper panel) and
NGC~4395 (lower panel) . For NGC~5457, circles represent arm
regions and triangles  inter-arm regions.}
\label{ewhar}
\end{figure}

In Figure \ref{ewhar} we show the behavior of the logarithm of H$\alpha$ equivalent width as a function of
galactocentric radius for NGC~5457 (upper panel) and  NGC~4395 (lower panel). For NGC~5457, arm regions are marked as
filled circles and inter-arm regions  as filled triangles.

\begin{figure}
\epsfig{file=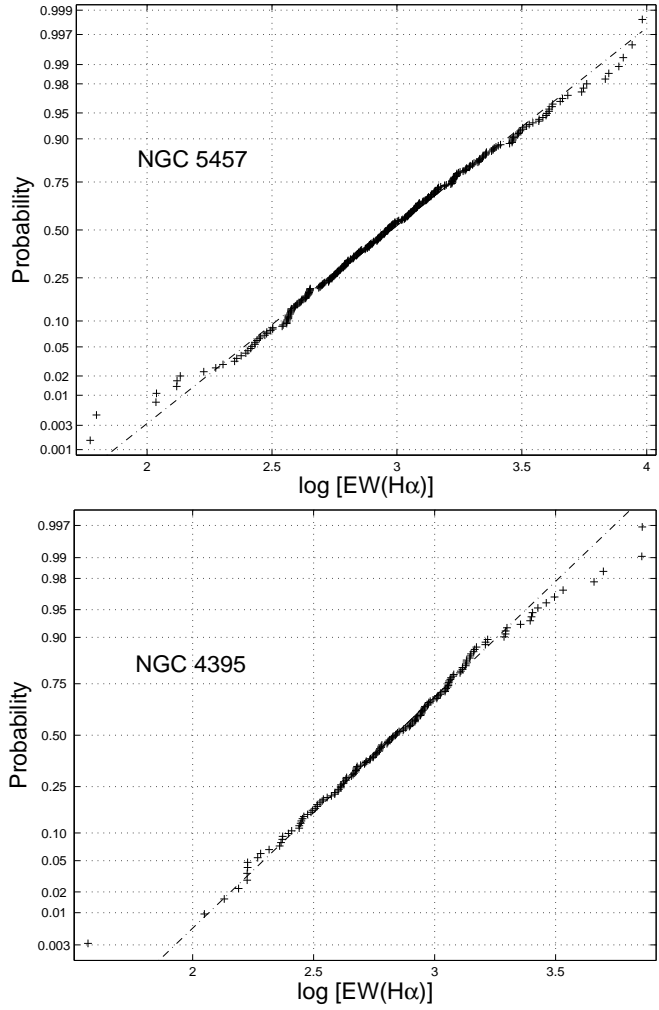,width=88mm}
\caption{Normal probability plot for the H$\alpha$ equivalent width for NGC~5457 (upper panel) and  NGC~4395 (lower panel). 
The crosses represent our data, and the
dashed line  the probability of an ``ideal'' normal distribution.}
\label{norm}
\end{figure}

The logarithm of the mean value for H$\alpha$ equivalent width is 2.97 $\pm$ 0.02 for NCG~5457 and  2.84 $\pm$ 0.03 for
NGC~4395. The error in these magnitudes was calculated as $\sigma/\sqrt{N}$, where $\sigma$ is the standard deviation 
($\sigma=0.37$ for NGC~5457 and $\sigma=0.35$ for NGC~4395)
and $N$ is the number of regions. To demostrate that 
the difference between both means is significant, two statistical tests were carried out:
 a two-sample test, applied to 
both distributions; and a $\chi^2$ test. 
 For the two sample test, we assume that both distributions are normal,  and that both have similar standard deviations. The 
normality of the
distributions can be tested with the normal probability plots in Figure~\ref{norm} for NGC~5457 (upper panel) and 
NGC~4395 (lower panel). If the data follow a normal distribution, the plot will be linear, which is indeed the case for both 
galaxies. The null hypothesis is that both 
means are equal. The results demostrate that the null hypothesis was false with  99.9\% confidence. In other words, 
the means were 
different and the value of the mean H$\alpha$ equivalent width for NGC~5456 was larger than that for NGC~4395. For the $\chi^2$ test, we obtained an $\chi^2=21.28$, and assuming 8 degrees of freedom (obtained as the number of bins of the distribution minus 1), this gives us that both distributions are different with a 99\% of confidence.
This
difference could be caused by four physical parameters: the age of the H~{\sc ii} regions 
(lower H$\alpha$ equivalent width with
age), the metallicity of the regions (lower metallicity implies higher equivalent width), 
 different H~{\sc ii} region leakage of ionizing photons (more leakage will reduce the H$\alpha$ equivalent width), and the
IMF of the regions (non-truncated IMFs show higher equivalent widths for low values of age), or a combination of all
or several of the aforementioned physical parameters. All these effects, except for the photon leakage, are
shown in Leitherer et al.'s (1999) models. The term ``truncated IMF'' is employed here with the same meaning as in Leitherer et al. (1999): a Salpeter IMF with a maximun mass for the stars of 30M$_{\odot}$, instead of the standard 100M$_{\odot}$ for the non-truncated Salpeter.

\begin{figure}
\epsfig{file=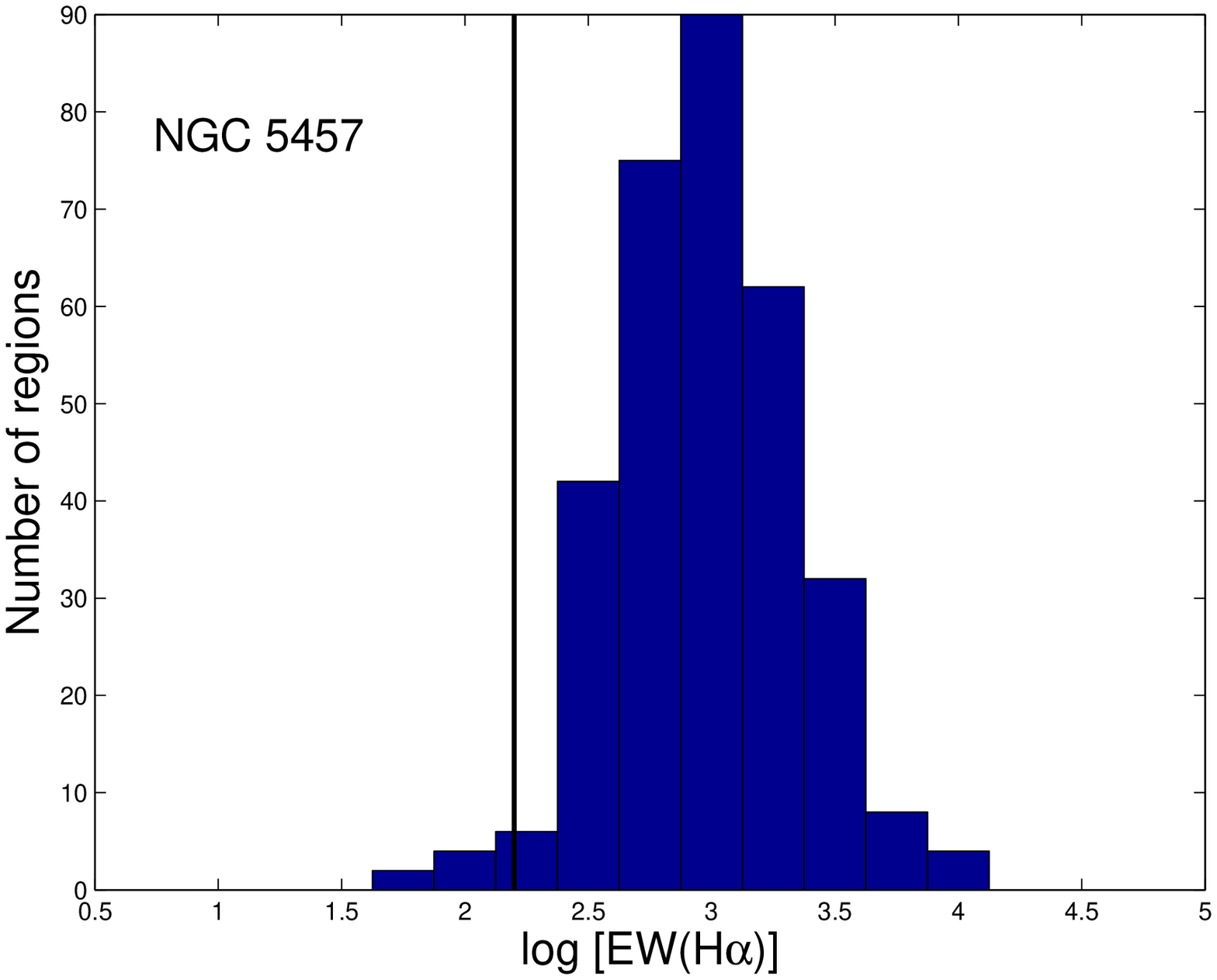,width=88mm}
\caption{Histogram of H$\alpha$ equivalent width for NGC~5457. The vertical line represents the limit where we begin to
fail to detect regions of lower
equivalent width.}
\label{hist54}
\end{figure}

\begin{figure}
\epsfig{file=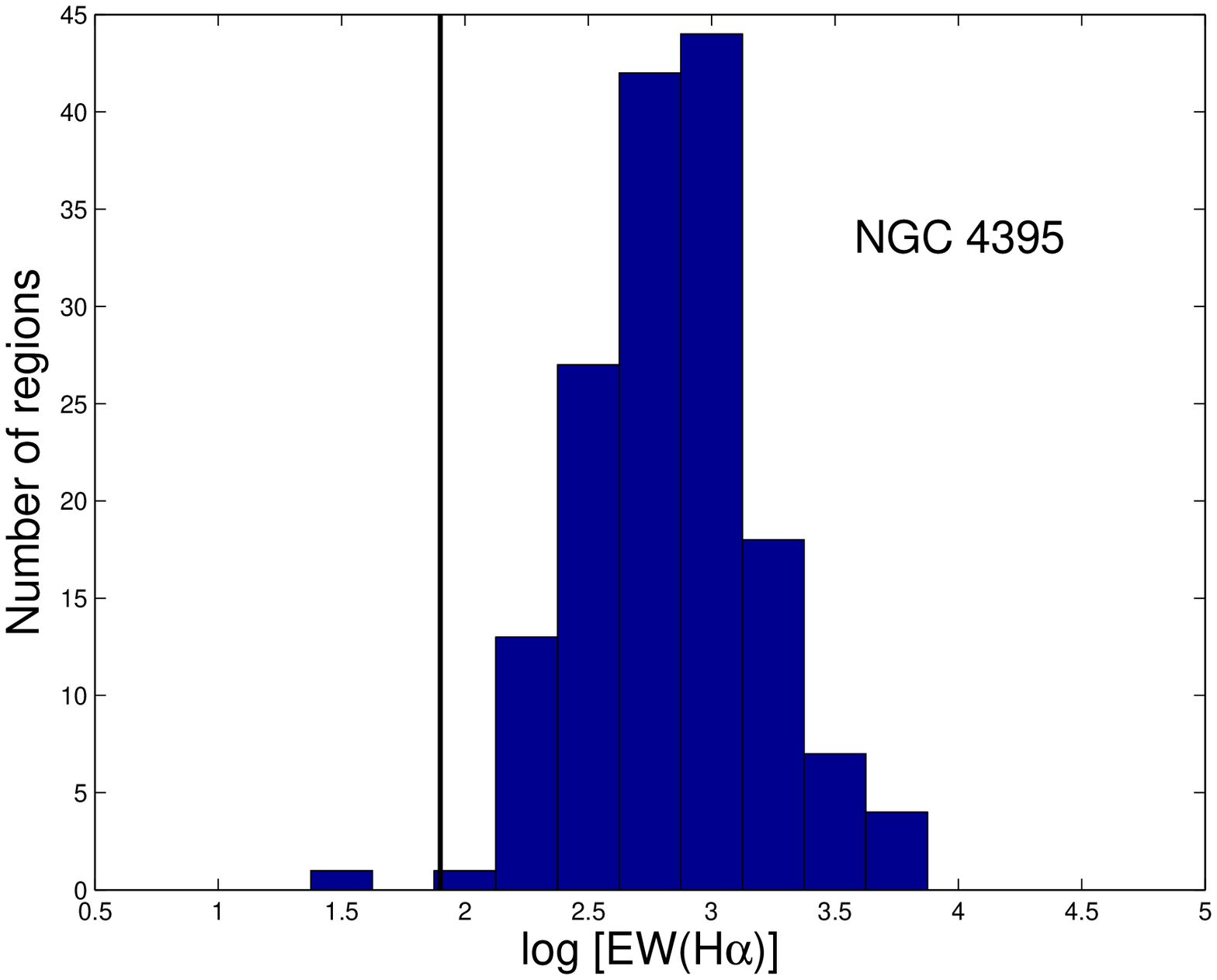,width=88mm}
\caption{Histogram of H$\alpha$ equivalent width for NGC~4395. The vertical line represents the limit where we begin to
fail to detect regions of lower
equivalent width.}
\label{hist43}
\end{figure}

Figures \ref{hist54} and \ref{hist43} are the histograms of the H$\alpha$ equivalent widths for NGC~5457 and NGC~4395
respectively. The vertical line in both figures indicates the zone where we begin to fail to 
detect regions with a low H$\alpha$
equivalent width. The value is $\log$[EW(H$\alpha$)] $\simeq$ 2.2 for NGC~5457 and $\log$[EW(H$\alpha$)] $\simeq$ 1.9 for
NGC~4395. These values where derived from the completeness limit of their respective luminosity functions, assuming a
mean value for the H$\alpha$ continuum flux. The completeness limit from the luminosity function was determined in the
bin with the largest number of H~{\sc ii} regions: $\log L_{\rm H\alpha}\simeq37.8$ for NGC~5457 and $\log
L_{\rm H\alpha}\simeq37$ for NGC~4395 where $L_{\rm H\alpha}$ is expressed in erg s$^{-1}$.

In Figure \ref{hist54} there are some H~{\sc ii} regions with  log\,EW(H$\alpha$) $\ge$ 3.5. The explanation for this
is that 
 these regions show a very large error because of weak H$\alpha$ continuum emission, and this generates 
 larger equivalent widths.
This effect, with the lack of H~{\sc ii} regions with low equivalent width, 
generates systematically higher values for the mean of
the H$\alpha$ equivalent width.

\section{Age effects}

\begin{figure}
\epsfig{file=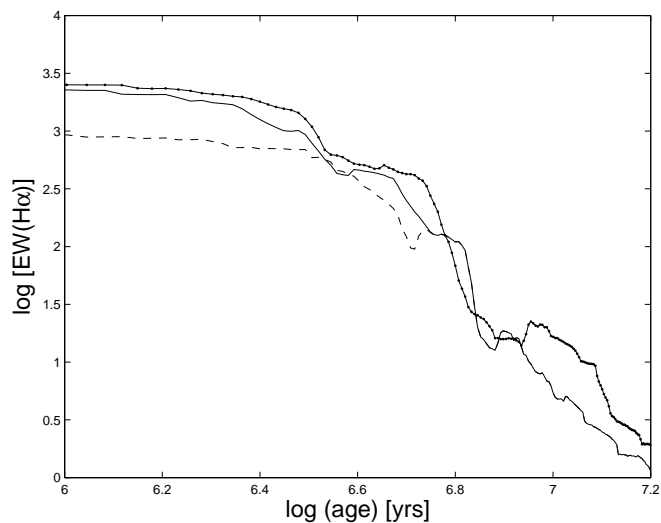,width=88mm}
\caption{H$\alpha$ equivalent width as a function of the age of regions for: a Salpeter IMF with a metalicity twice the solar value 
(continuous line); a truncated Salpeter IMF with a metallicity twice the solar value (dashed line), and a Salpeter IMF with a
 metalicity equal to the solar value (line with dots). Extracted from Leitherer et al.\ (1999) models.}
\label{leit}
\end{figure}

From the models presented in Leitherer et al\. (1999), it is clear that the stronger dependence of the H$\alpha$ equivalent width 
comes from the age of the H~{\sc ii} regions. In Figure~\ref{leit} we have represented several of the models from Leitherer 
et al.\ (1999). If  age is the main cause of the observed difference, the H~{\sc ii} regions of NGC~5457 should be younger than 
the H~{\sc ii} regions of NGC~4395. However, if we consider the whole population of  H~{\sc ii} regions for the two galaxies, 
taking into account the short 
lifetime of an H~{\sc ii} region (about 6 to 10 Myr for the oldest regions [Coppetti et al.\ 1986]) and the large number of H~{\sc ii} 
regions, it is reasonable to assume that the mean age for the
H~{\sc ii} regions of both
galaxies should be similar. Moreover, even if a star-forming front exists in the arms of the galaxy because of 
the  triggering of
star formation by any of the processes described in Elmegreen (1992), these H~{\sc ii} regions will
show a full range of ages when considering arm and interarm regions. Moreover, none of the studied galaxies are strong starbust or poststarbust.

\begin{figure}
\epsfig{file=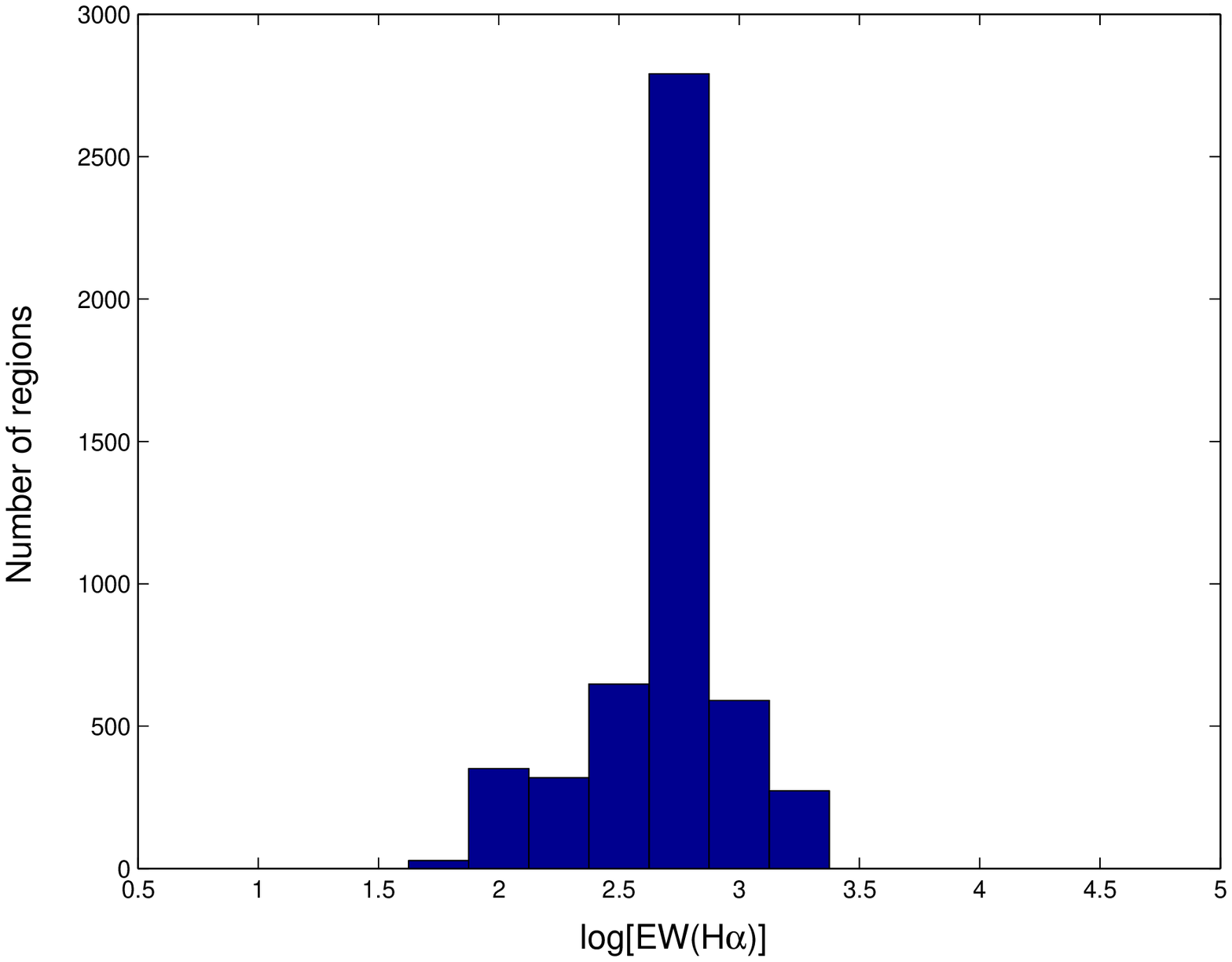,width=88mm}
\caption{Histogram of the H$\alpha$ equivalent width, obtained through Leitherer et al.\ (1999) models for a Salpeter IMF and solar 
metallicity assuming that there is no stellar formation in the last 3 Myr.}
\label{starfr}
\end{figure}

 The possibility of a difference in H$\alpha$ equivalent width caused by
 a difference in the star formation rate (SFR) also exists. In fact, a gross estimate
  of the SFR in both galaxies using the H$\alpha$ luminosity (Kennicutt 1983) shows  that the SFR per  unit area
  for the grand design galaxy is three times larger than the SFR for the flocculent galaxy. Taking into account 
  a very simple model that generates H~{\sc ii} regions with different SFRs
   but with the same IMF,
    and that then tranforms age into H$\alpha$ equivalent width through Leitherer et al.\ (1999) models, 
    the resultant equivalent widths are too close to be distinguished in these observations. Taking into account that significant SFR
     variations occur over large periods of time (Kennicutt et al.\ 1994) and the short lifespan of the H~{\sc ii} regions, we have 
     assumed that the the SFR was constant for at least 3\,Myr. 
    To obtain a difference equivalent to that observed  it is necessary that star-forming processes 
    in the flocculent galaxy should not have occurred
     in the last 3\,Myr, which does not agree with the observed data for NGC~4395. Moreover, is impossible to simulate the shape of 
     the distribution (see Figure~\ref{starfr}).
On the other hand, the star formation history (SFH) of both galaxies could be very different in a cosmological time scale. But this difference would rather show in the continuum than in emission line data.Here we deal with H~{\sc ii} regions that have a time scale from 1 to 10Myr (approximately). Moreover, although it is true that both galaxies present a very different SFR, such different histories do not affect the H$\alpha$ equivalent width as much as the IMF. Due to the short lifespan of H~{\sc ii} regions, the only remanent of a very different SFH for a large periods of time would be old low-intermediate mass stars. These stars will only make a contribution to the continuum of the disk, and such subyacent emission has been wiped out as shown in Cedr\'es \& Cepa (2002).
 \\

There is a relation between Hubble type and the H$\alpha$ equivalent witdh (see for example, Gavazzi et al.\ 2002, and references therein). 
However, in this relation the Sm galaxies present a larger H$\alpha$ equivalent width than the Sc--Sd galaxies, contrary to our results, 
where NGC~5457, an SAB(rs)cd type galaxy (de Vaucouleurs et al.\ 1991) has a larger value for the equivalent width than NGC~4395, 
an SA(s)m type galaxy (de 
Vaucouleurs et al.\ 1991).

\section{Abundance effects}

The Leitherer et al.\ (1999) models clearly indicate a relation between the metalicity of the regions and the H$\alpha$ equivalent width. 
However, this relation is not as strong as the age or the IMF (see Figure~\ref{leit}). The metallicity for the inner parts of NGC~5457 
is higher than that for NGC~4395 (Cedr\'es \& Cepa 2002), so, in agreement  with the  Leitherer et al.\ (1999) models, the mean 
equivalent width for NGC~4395 should be almost equal to or greater  (100--200 \AA) than the mean equivalent width of NGC~5457, 
contrary to the results obtained here. The difference observed would then be even larger when taking metallicity into account.

\section{Monte Carlo tests}
In order to explore the IMF and photon leakage of other possible factors in the detected difference in the H$\alpha$ equivalent width,
 several Monte Carlo tests were computed.

The random variable chosen to develop the test was the age of the H~{\sc ii} regions. We assumed a continuum 
distribution of age between 1\ Myr and 
7\ Myr as maximum, on the assumption that  there is no privileged age, so all the ages of the H~{\sc ii} regions 
have the same probability of existence at 
the present time between the detection boundaries. However, in order to adjust the results to the observations to a better extend, for several simulations the maximum age was changed to aproximately 6.8\ Myr. The total number of simulated regions was 10\,000. This number was selected because 
it is large enough to avoid statistical uncertainties with reasonable computation time. A random value for the error (between 0.1 to 0.35 
dex in H$\alpha$ equivalent width) was introduced for each simulated region, in order to reproduce to some extend the errors presented 
in the data. 

To obtain the values of the equivalent width, we employed Leitherer et al.\ (1999) models, assuming an intantaneous burst, and transforming age directly into equivalent width.

Taking into account that NGC~5457 has a steep metallicity gradient (van Zee et al.\ 1998 and references therein), 
and that in its inner zones 
it is about $Z=2Z_{\odot}$ and $Z=Z_{\odot}$ at $R=6^{\prime}$, it is necessary to employ at least two different metallicities to 
simulate this behavior. However, recent measurements of the abundance for this galaxy (Kennicutt et al.\ 2003, Cedr\'es et al.\ 2004) 
show that the metallicity could be lower than $Z_{\odot}$, so an abundance of $Z=0.4Z_{\odot}$ was also employed in our simulations.  
This method was developed because it allows us to use the Leitherer et al.\ (1999) tabulated data directly without 
interpolation.

\begin{figure}
\epsfig{file=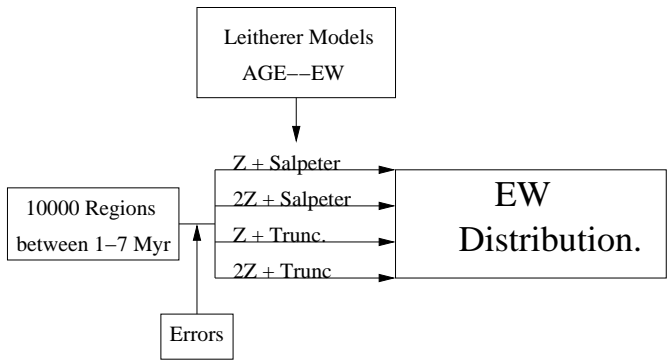,width=88mm}
\caption{Flux diagram of the Monte Carlo simulation for the H$\alpha$ equivalent width. The input for the simulation are the 10000 regions randomly generated between 1 to 7 Myr. For each age, an error is introduced to simulate the observational uncertainties. After that, employing Leitherer et al. (1999) Age--EW relationship, and using different combinations of metallicity and IMF, the age distribution is transformed into an EW distribution.}
\label{diag1}
\end{figure}

 In Figure \ref{diag1} we show the flux diagram of a summary of a simulation. In this process, after the regions are created, for each one an individual error is introduced to simulate the uncertainties of our observed data. After that, we use the age-generated regions as the input for Leitherer et al. (1999) models, with different metallicites and IMF. The final output of the models is an equivalent width distribution which includes all the possible effects due to variations of the age of the regions, different IMF, the metallicity gradiend and observational uncertainties.

\begin{figure}
\epsfig{file=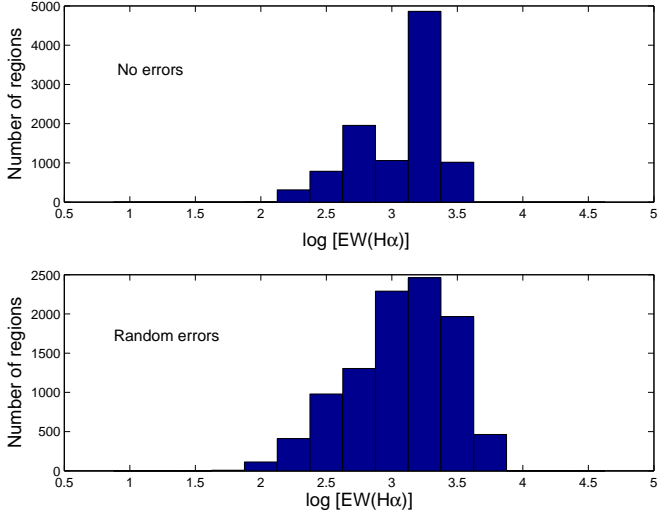,width=88mm}
\caption{Monte Carlo simulations for a Salpeter IMF with two metallicities: solar and twice solar. Upper panel: regions without error. 
Lower panel: regions with random error.}
\label{error}
\end{figure}

In Figure \ref{error} we represent the simulated results for 10\,000 regions, half of them with twice the solar metallicity and the 
other half with solar metallicity, and a Salpeter IMF without errors (upper panel) and with the random errors (lower panel).

\section{IMF effects}

From the Leitherer et al.\ (1999) models it is clear that there is a strong dependence of the H$\alpha$ equivalent width and the IMF for 
H~{\sc ii} regions with younger ages (see Figure \ref{leit}). So it is reasonable to assume that a difference in the IMF between 
both galaxies could explain the different H$\alpha$ equivalent width observed.
Employing the Monte Carlo tests described in the previous section, we developed
several models with different IMFs  to explore 
this possibility.

\begin{figure}
\epsfig{file=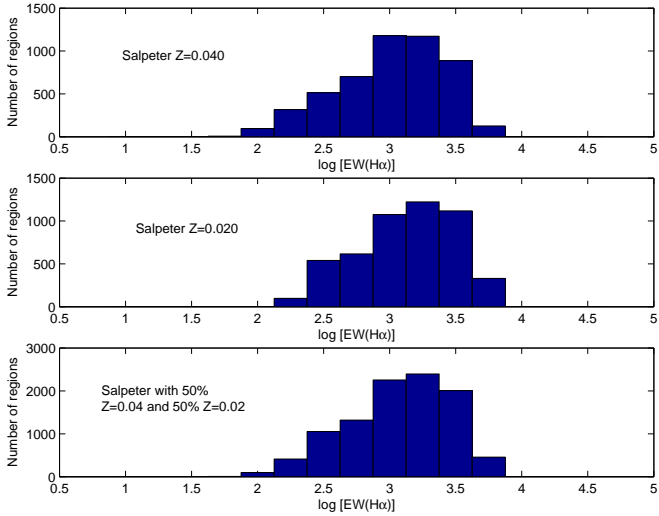,width=88mm}
\caption{Monte Carlo results for a Salpeter IMF and two metallicities. Upper panel: $Z=2Z_{\odot}$. Middle panel:  
$Z=Z_{\odot}$. Lower panel: sum of middle and upper panels.}
\label{metalo}
\end{figure}

In Figure \ref{metalo} we present the Monte Carlo results for a single Salpeter IMF and two metalicities. It is clear that, even 
considering two different metallicities, it is not possible to reproduce the observed 
trend for the H$\alpha$ 
equivalent width in the grand design galaxy with sufficient accuracy. It is therefore necessary to assume that a certain 
number of H~{\sc ii} 
regions may have another kind of IMF.

\begin{figure}
\epsfig{file=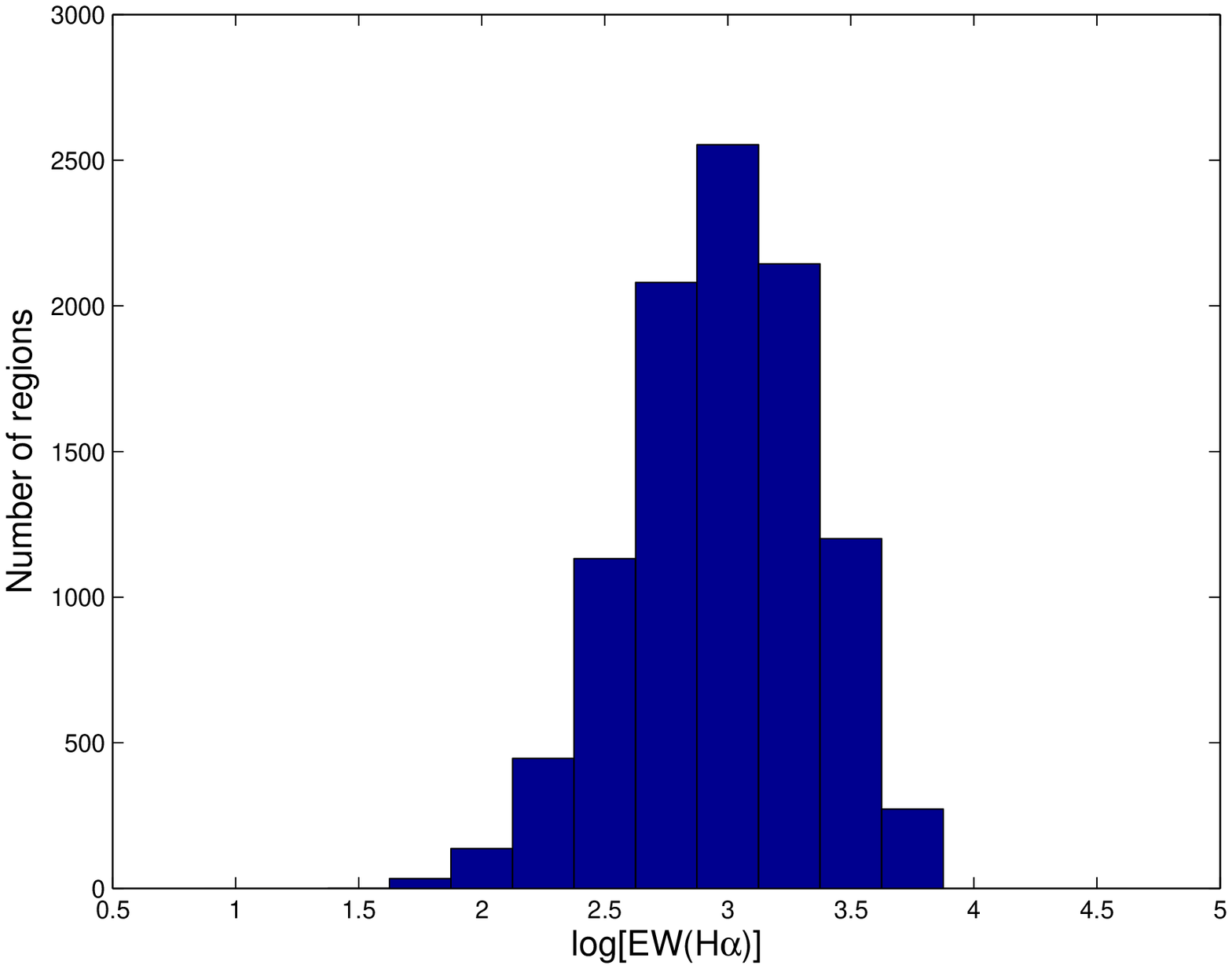,width=88mm}
\caption{Best Monte Carlo simulation for the galaxy NGC~5457. This model has 62\% of its regions with a Salpeter IMF and 38\% of its
regions with a truncated Salpeter IMF. The maximum age is 6.75 on a logartithmic scale and half of the regions have 
twice the solar metallicity. 
The other half have  solar metallicity. This model is marked in Table~\ref{para1} as number 4.}
\label{bestgd}
\end{figure}

In Table \ref{para1} a sample of the  parameters employed for several simulations for the grand design galaxy is provided, with the best 
result marked in boldface. The best simulation was obtained with  62\% of regions using a Salpeter IMF and  40\% of regions using a 
truncated Salpeter IMF. For this simulation, we assumed that 50\% of the regions had  solar metallicity and the other 50\% twice 
 solar. The maximum age was chosen to be 6.75 on a logarithmic scale. This model is represented in Figure~\ref{bestgd}.

\begin{table}
\begin{tabular}{c|c|c|c|c|c|c|c|c|c}
\hline
\multicolumn{2}{c|}{IMF} &\multicolumn{3}{|c|}{Met.} & Max.\ Age & 2 TEST & Mean & STD & Number\\
Sal. & Trunc. & 2$Z_\odot$ & $Z_\odot$ & 0.4$Z_\odot$ & & & & &\\ 
\hline
50\% & 50\% & 50\% & 50\% & 0\% & 6.8 & $<$1\% & 2.89 & N.A.& 1\\
50\% & 50\% & 50\% & 50\% & 0\% & 6.75 & 19\% & 2.94 & N.A. & 2\\
60\% & 40\% & 50\% & 50\% & 0\% & 6.75 & 98.7\% & 2.97 & 0.37& 3\\
62\% & 38\% & 50\% & 50\% & 0\% & 6.75 & \bf{99\%} & 2.97 & 0.375 & 4\\
70\% & 30\% & 50\% & 50\% & 0\% & 6.8 & 10\% & 2.94 & N.A.& 5\\
70\% & 30\% & 50\% & 50\% & 0\% & 6.75 & 30\% & 2.99 & N.A.& 6\\
50\% & 50\% & 40\% & 50\% & 10\% & 6.75 & 60\% & 2.96 & N.A.& 7\\
60\% & 40\% & 50\% & 40\% & 10\% & 6.75 & 87\% & 2.97 & N.A.& 8\\
60\% & 40\% & 50\% & 30\% & 20\% & 6.75 & 68\% & 2.96 & N.A.& 9\\
60\% & 40\% & 40\% & 50\% & 10\% & 6.75 & 64\% & 2.98 & N.A.& 10\\
60\% & 40\% & 40\% & 50\% & 10\% & 6.8 & 5\% & 2.92 & N.A. & 11\\
60\% & 40\% & 30\% & 50\% & 20\% & 6.75 & 30\% & 2.99 & N.A. & 12\\
\hline
\end{tabular}
\caption{Parameters of the Monte Carlo simulations for NGC~5457. The  first two columns are the percentage of the regions with a 
determinate IMF. The first column is for a Salpeter IMF. The second column is for a truncated Salpeter IMF. Columns 3 to 5 are the
 percentage of the regions with a determinate metallicity. The column 3 is the percentage of regions with a metallicity twice  
 solar. Column 4 is the percentage of regions with a solar metallicity. Column 5 is the percentage of 
 regions with a metallicity 0.4 times  solar. Column 6 is the maximum age of the regions created on a logarithmic scale. Column 7 
 is the confidence of the two sample test when comparing the model with the data. The higher the percentage, the more similar the 
 simulation is to  real data. Column 8 is the mean value of the distribution, and column 9 is the standard deviation. This last 
 parameter is presented only for the best simulations to the data. Column 10 is the identification number of the model.}
\label{para1}
\end{table}

The best result obtained (highlighted in Table \ref{para1}) is showed in Figure~\ref{bestgd}. Compared with the data represented in
 Figure~\ref{hist54}, it is clear that there is a very good agreement between the simulated data and the observed data, not only in 
 the shape of the distribution but in the mean value and in the standard deviation of the data.

\begin{figure}
\epsfig{file=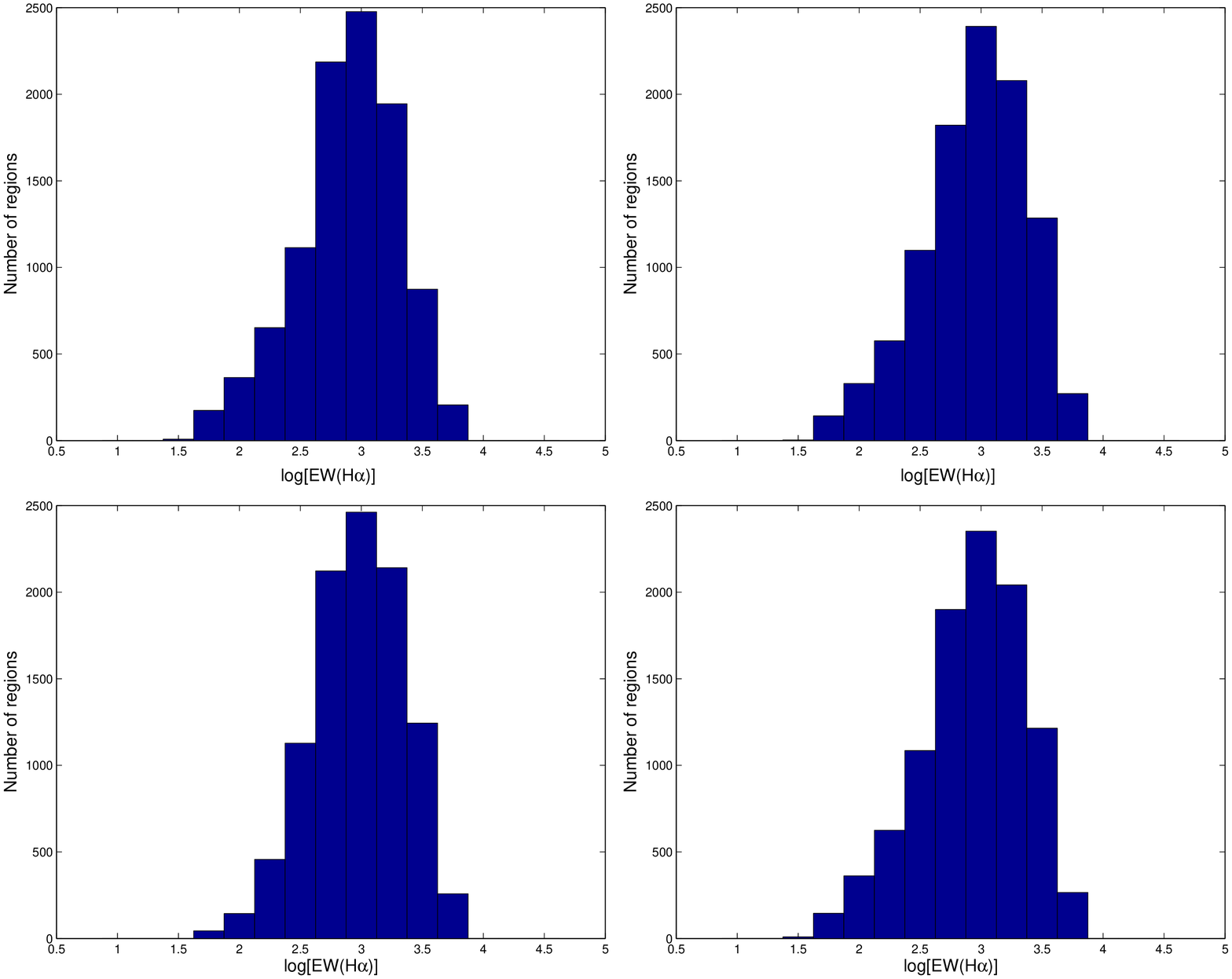,width=176mm}
\caption{Monte Carlo models for NGC~5457. Models are: top left, model 1; top right, model 5; bottom left, model 8; bottom right, model 11.}
\label{mosgd1}
\end{figure}

In Figure \ref{mosgd1}, we represent some of the models summarized in Table~\ref{para1}. The models are from left to right and top to bottom,  models 1, 5, 8, and 11 respectively. From Table~\ref{para1}, and taking into account the 
results of the statistical test carried out (the two sample test), it is clear that the best coincidence between the models and the observed 
data is obtained when it is assumed that the majority of the regions present a Salpeter IMF, even when considering two or three different 
metalicities. Moreover, the best results are also obtained assuming that the oldest regions have $\log ({\rm age})=6.75$, instead of 
$\log ({\rm age})=6.8$.

For the flocculent galaxy (NGC~4395), there is no trace of a metallicity gradient (or if there is it
 is very shallow [Roy et al.\ 1996; Van Zee et al.\ 1998]). However, the large dispersion of the data for this galaxy (Cedr\'es \& 
 Cepa 2002) suggests that there is the possible coexistence of two different metallicities in order to simulate this parameter.  
The parameters employed for several simulations are summarized in Table~\ref{para2}. The best results are highlighted in boldface.

\begin{table}
\begin{tabular}{c|c|c|c|c|c|c|c|c}
\hline
\multicolumn{2}{c|}{IMF} &\multicolumn{2}{|c|}{Met.} & Max. Age & 2 TEST & Mean & STD & Number\\
Sal. & Trunc. & 2$Z_\odot$ & $Z_\odot$ & & & & &\\ 
\hline
40\% & 60\% & 50\% & 50\% & 6.8 & $<$ 0.1\% & 2.87 & N.A. & 1\\
40\% & 60\% & 50\% & 50\% & 6.75 & 1\% & 2.92 & N.A. & 2 \\
20\% & 80\% & 0\% & 100\% & 6.8 & 7\% & 2.89 & N.A. & 3\\
0\% & 100\% & 0\% & 100\% & 6.8 & 39\% & 2.85 & N.A. & 4\\
0\% & 100\% & 10\% & 90\% & 6.8 & {\bf 90\%} & 2.83 & 0.33 & 5 \\
0\% & 100\% & 45\% & 55\% & 6.75 & {\bf 97\%} & 2.837 & 0.29 & 6\\
\hline
\end{tabular}
\caption{Parameters of the Monte Carlo simulations for NGC~4395. The two first columns are the percentage of the regions with a 
determinate IMF. The first column is for a Salpeter IMF. The second column is for a truncated Salpeter IMF. Columns 3 and 4 are the
 percentage of the regions with a determinate metallicity. The column 3 is the percentage of regions with a metallicity twice  
 solar. Column 4 is the percentage of regions with a metallicity equal to the solar value. Column 5 is the maximum age of the 
 regions created on a logarithmic scale. Column 6 is the confidence of the two sample test when comparing 
 the model with the data. The higher 
 the percentage, the more similar the simulation is to the real data. Column 7 is the mean value of the distribution, and column 8 is
  the standard deviation. This last parameter is presented only for the best simulations to the data. Column 9 is the identification 
  number of the model}
\label{para2}
\end{table}

\begin{figure}
\epsfig{file=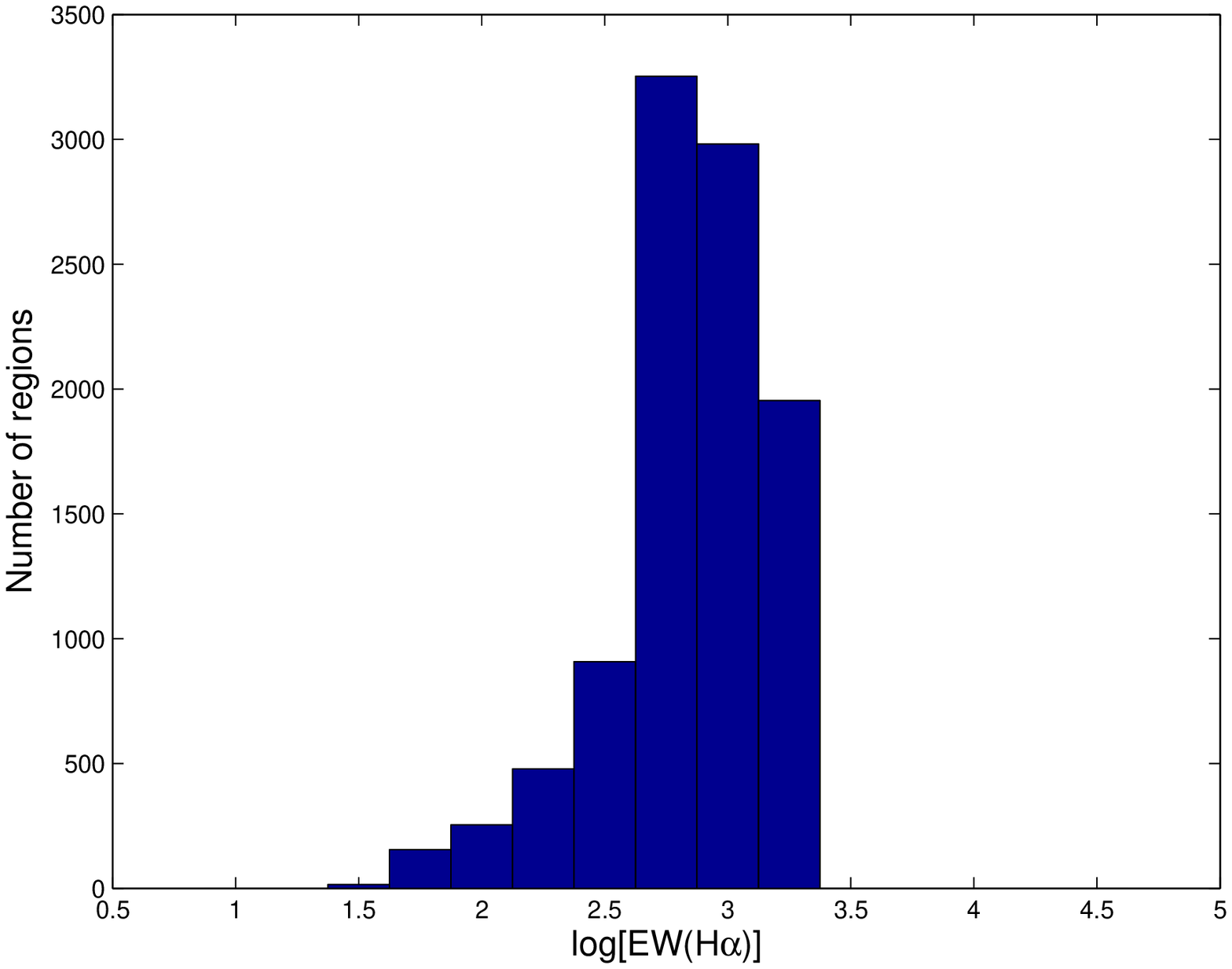,width=88mm}
\caption{Monte Carlo simulation for NGC~4395. In this simulation,  100\% of the regions have a truncated Salpeter IMF. The maximum 
age is 6.8 on a logarithmic scale. Ten percent of the regions have a metallicity twice  solar  and the remaining 90\% 
have  solar metallicity.
 This model is marked with the number 5 in Table~\ref{para2}.}
\label{best1f}
\end{figure}

\begin{figure}
\epsfig{file=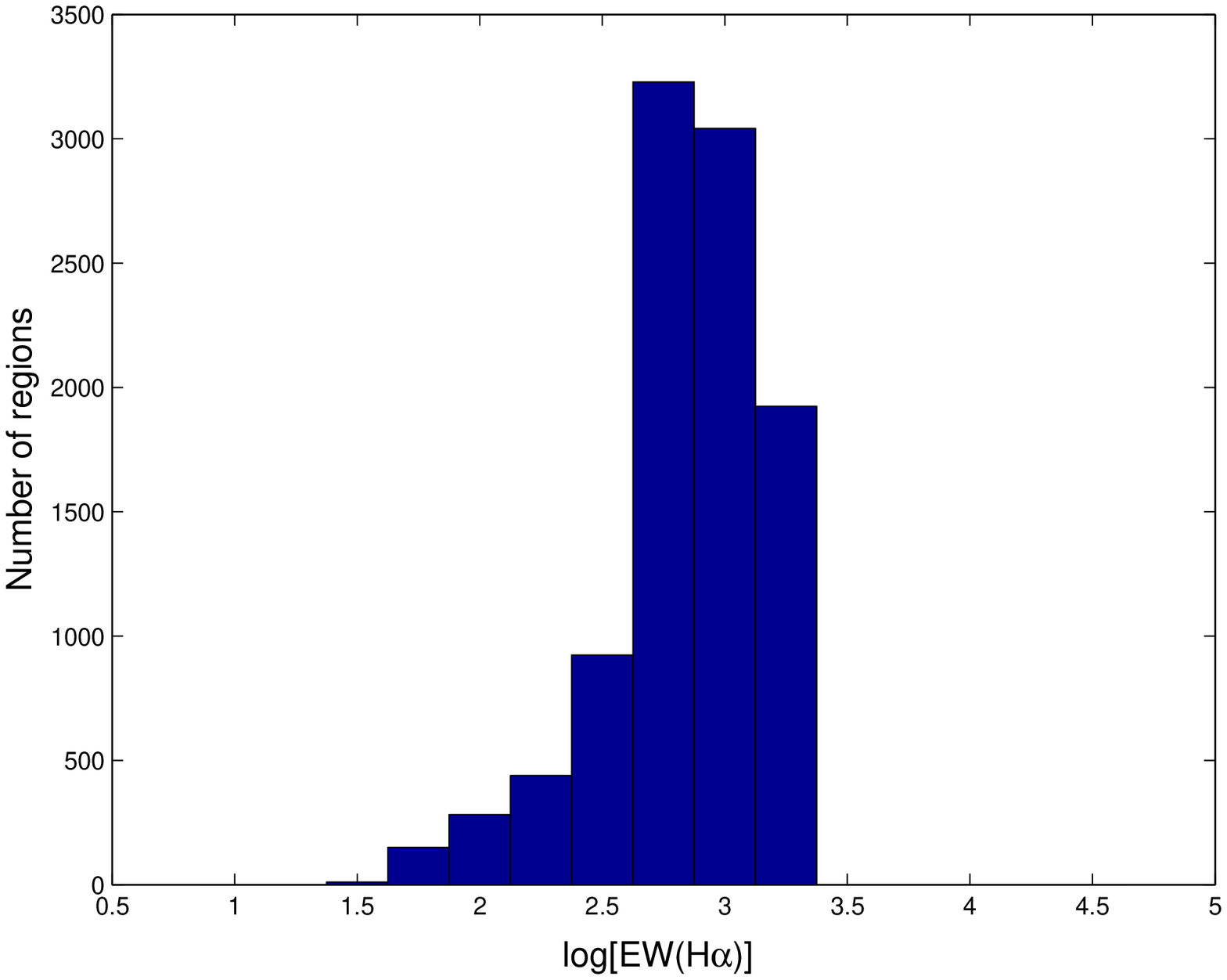,width=88mm}
\caption{Monte Carlo simulation for NGC~4395. In this simulation,  100\% of the regions have a truncated Salpeter IMF. The maximum 
age is 6.75 on a logarithmic scale. Forty-five percent of the regions have a metallicity twice  solar  and the remaining 55\% have  solar 
metallicity. This model is marked with the number 6 in Table~\ref{para2}.}
\label{best2f}
\end{figure}

In Figures \ref{best1f} and \ref{best2f} we have represented the models 5 and 6. The model in Figure \ref{best1f} has 100\% of its regions 
with a truncated Salpeter IMF, 10\% of its regions with a metallicity twice  solar,  and the remaining 90\% of its regions with  solar 
metallicity. The maximum age was 6.8 on  a logarithmic scale. 
The mean value for the logarithm of the H$\alpha$ equivalent width is 2.83 and the 
standart deviation is 0.33. Both numbers are very close to those obtained from the observed data. However, the shape of the distribution 
it is not as accurate as for the grand design galaxy, although the confidence of the two-sample test is 90\%. For Figure~\ref{best2f},
 we obtain a better coincidence with 55\% of the regions with solar metallicity and 45\% of the regions with a metallicity twice  solar 
  and with a maximum age of 6.75 on  a logarithmic scale. 
  However,  there is again a problem with the shape of the distribution. As seen in 
 the two previous examples, it is difficult to obtain for this galaxy a unique set of parameters to simulate the distribution of 
 equivalent widths. But the only way to approximate to a high value for the confidence in the two sample test is by considering  all 
 the regions in this galaxy to present a truncated IMF.

For this galaxy, the majority of the models show a maximum age of 6.8 on a logarithmic scale. This can be explained in terms of the galaxy being  nearer than the grand design galaxy, so we are able to observe fainter regions in H$\alpha$, that generally are the older ones.

\section{Photon leakage effects}

Photon leakage could be a possible cause of the observed H$\alpha$ equivalent width distribution. According to Zurita 
et al.\ (2002), the more luminous the region is, the larger is the fraction of escaping photons. In Figure~\ref{almu} we present two models 
from Zurita et al.\ (2002), one with a constant filling factor for the H~{\sc ii} regions (circles) and one with a constant density 
(crosses). In order to simulate the photon leakage employing the models of Starbust'99, it is necessary to rescale the 
relationship presented there between the ionizing photons and the age of the H~{\sc ii} regions (figure 77b of Leitherer et 
al.\ 1999). The models there have been normalized to a total mass of $10^6\ M_{\odot}$ per H~{\sc ii} region, so the number of 
ionizing photons is far too large. Kennicutt's (1988) expression for the total stellar ionizing mass in the range 
$10\ M_{\odot}<M<100\ M_{\odot}$, with a Salpeter IMF is:
\begin{equation}
M_s=3.6\times10^{-36}L_{\rm H\alpha},
\end{equation}
where $M_s$ is in solar masses and $L_{\rm H\alpha}$ is in erg\,s$^{-1}$.

We can relate the H$\alpha$ luminosity with the number of ionizing photons according to Osterbrock (1989):
\begin{equation}
Q(H^0)=7.31\times10^{11}L_{\rm H\alpha},
\end{equation}
where $L_{H\alpha}$ is in erg\,s$^{-1}$.

So comparing the distribution of ionizing masses between our data from NCG~5457 and a simulated distribution of ionizing masses 
employing the Leitherer et al.\ (1999) models, and assuming a Salpeter IMF and a similar range of ages for the H~{\sc ii} regions, we 
obtained a correction factor of $(2.95\pm0.1)\times10^{-3}$. The uncertainties in this quantity were calculated by varying the value
 of the factor and observing the variation in the result of the final simulation.
 
Employing the relationship between $Q(H^0)$ and the rescaled age of the H~{\sc ii} region, we obtained a relation between $Q(H^0)$ 
and the H$\alpha$ equivalent width through the models of Leitherer et al.\ (1999). With these two relationships, we were able to run 
Monte Carlo tests as described in section 6. In this case, because of the large number of parameters employed and in order
to avoid  excessive
noise in the results, each Monte Carlo test with 10\,000 simulated regions was repeated 1000 times.

\begin{figure}
\epsfig{file=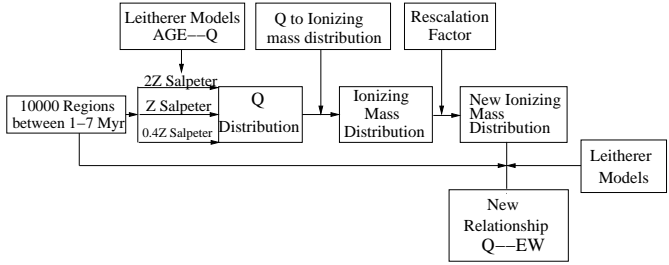,width=88mm}
\caption{Summary of the process to obtain a corrected relationship for our galaxies between the number of photons (Q) and the H$\alpha$ equivalent width. As in Figure \ref{diag1}, the input is the age distribution for 10000 regions, but in this case, the regions are transformed, employing Leitherer et al. (1999) relationship between age and Q into a photon distribution with a Salpeter IMF and several metallicities. The process then transforms Q into an ionizing mass distribution. This distribution is rescalated to adjust it to the galaxy H~{\sc ii} ionizing mass. Employing again Leitherer et al. (1999) models, a valid relationship for our galaxy between the number of ionizing photons and the H$\alpha$ equivalent width is generated.}
\label{diag2}
\end{figure}

\begin{figure}
\epsfig{file=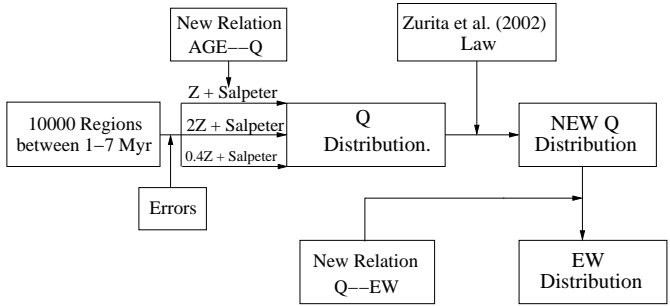,width=88mm}
\caption{Flux diagram for the Monte Carlo simulation for the H$\alpha$ equivalent width with photon leaking. For this case, the input and errors are introduced as explained in Figure \ref{diag1}. Employing the new relationship between age and Q, obtained as showed in Figure \ref{diag2}, we achieved the Q distribution. We applied Zurita et al. (2002) law to this distribution to supress the photons that may leak for the H~{\sc ii} regions, getting a new Q distribution. Employing the new relationship between Q and the equivalent width (indicated in Figure \ref{diag2}), we obtained the definitive H$\alpha$ equivalent width distribution with photon leakage.}
\label{diag3}
\end{figure}

\begin{figure}
\epsfig{file=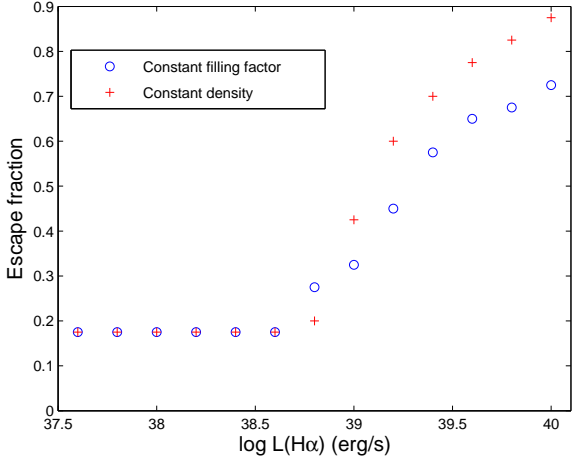,width=88mm}
\caption{Fraction of the total ionizing photon flux within an H~{\sc ii} region  escaping from it according to 
Zurita et al.'s (2002) models. The circles are from a model with constant a filling factor and the crosses from a model with 
constant density.}
\label{almu}
\end{figure}

In Figures \ref{diag2} and \ref{diag3} we represent the flux diagram of the process to obtain a corrected relationship between the number of photons and the equivalent width, and the new Monte Carlo simulation, respectively.

\begin{table}
\begin{tabular}{c|c|c|c|c|c|c}
\hline
\multicolumn{3}{c|}{Met.} & Escape & 2 TEST & Mean & Number\\
    2$Z_\odot$ & $Z_\odot$ & 0.4$Z_\odot$ & Law  & & & \\ 
\hline
 50\% & 50\% & 0\% & den. & $<$1\% & 2.93$\pm$0.01 & 1\\
 50\% & 50\% & 0\% & f.f. & $<$1\% & 2.92$\pm$0.01 & 2\\
  0\% & 100\% & 0\% & den. & 78\% & 2.97$\pm$0.01 & 3\\
  0\% & 100\% & 0\% & f.f. & 72\% & 2.97$\pm$0.01 & 4\\
  0\% & 92.5\% & 7.5\% & den. & 81\% & 2.97$\pm$0.01 & 5\\
  0\% & 92.5\% & 7.5\% & f.f. & 79\% & 2.97$\pm$0.01 & 6\\ 
\hline
\end{tabular}
\caption{Parameters of the Monte Carlo simulations for NGC~5457 with photon leakage. Columns 1 to 3 are the percentage of  
regions with a determinate metallicity. Column 1 is the percentage of regions with a metallicity twice  solar. 
Column 2 is the percentage of regions with a metallicity equal to the solar value. Column 3 is the percentage of regions with a 
metallicity 0.4 times the solar value. Column 4 is the photon escape law employed from Zurita et al.\ (2002). ''Den'' indicates 
constant density and ``f.f.'' indicates constant filling factor. Column 5 is the confidence of the two sample test when comparing the 
model with the data. The higher the percentage, the more similar the simulation to the real data. Column 6 is the mean value of the
 distribution. Column 7 is the identification number of the model. A Salpeter IMF was employed for all the models.}
\label{escgd}
\end{table}

\begin{figure}
\epsfig{file=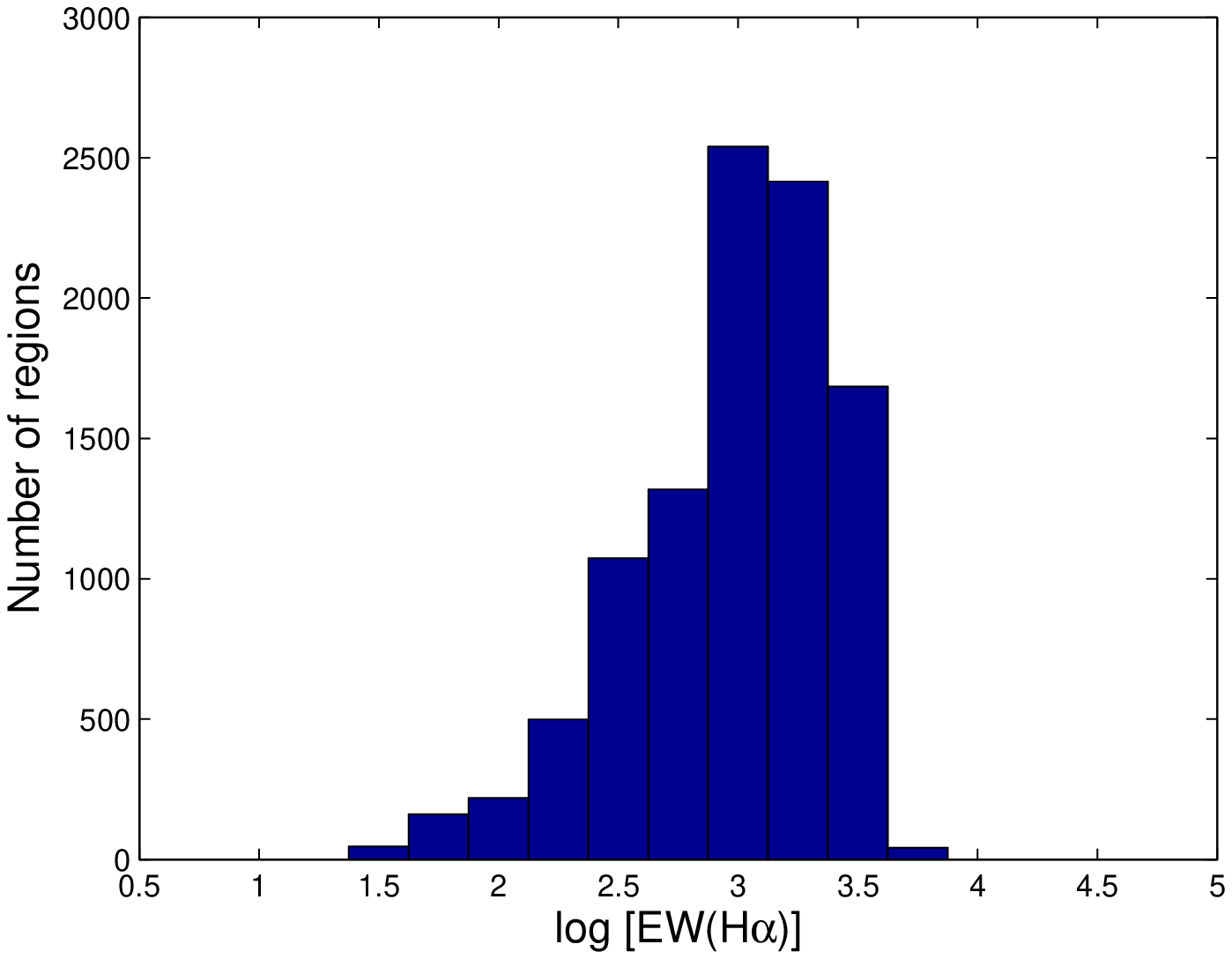,width=88mm}
\caption{One of the Monte Carlo simulations. In this simulation 100\% of the regions have a Salpeter IMF.  92.5\% of the regions 
have a metallicity equal to the solar value and the remaining 7.5\% have a metallicity 0.4 times the solar. This simulation is part 
of the models marked with number 5 in Table \ref{escgd}.}
\label{histesm}
\end{figure}

In Table \ref{escgd} we show the results of the simulations for the grand design galaxy. The column distribution is the same as in
 Table~\ref{para1}, but in this case we have introduced a new column that indicates the law of photon escape employed (``den'' 
 for constant density and ``f.f.'' for constant filling factor). The column for the maximun age of the regions was suppressed because 
 it was selected as $\log({\rm age})=6.8$ for all the simulations. The column for the IMF was also suppressed because 
 A Salpeter IMF was employed 
  for all the simulations. Simulations with a maximun age of $\log({\rm age})=6.75$ were carried out, but showed no significant 
 variation when compared with $\log({\rm age})=6.8$.
From Table~\ref{escgd} it is clear that there is no large difference between the simulations with a photon escape law with constant
 density and a photon escape law with constant filling factor; however, the results for an escape law with constant density are always 
 better matched to real data than those for a law with a constant filling factor.
In general, the simulations do not get so close to the real dara as the simulations with mixed IMFs and no photon escape. However, 
simulation number 5 in Table~\ref{escgd} has a good level of similarity with the real data. This simulation presents a peculiar
 metallicity mixture that  is compatible with a less metallic NGC~5457, as proposed in Kennicutt et al.\ (2003) and 
 Cedr\'es et al.\ (2004).  The histogram of one of the simulations obtained employing the 
 parameters in number 5 is represented in Figure~\ref{histesm} .

\begin{figure}
\epsfig{file=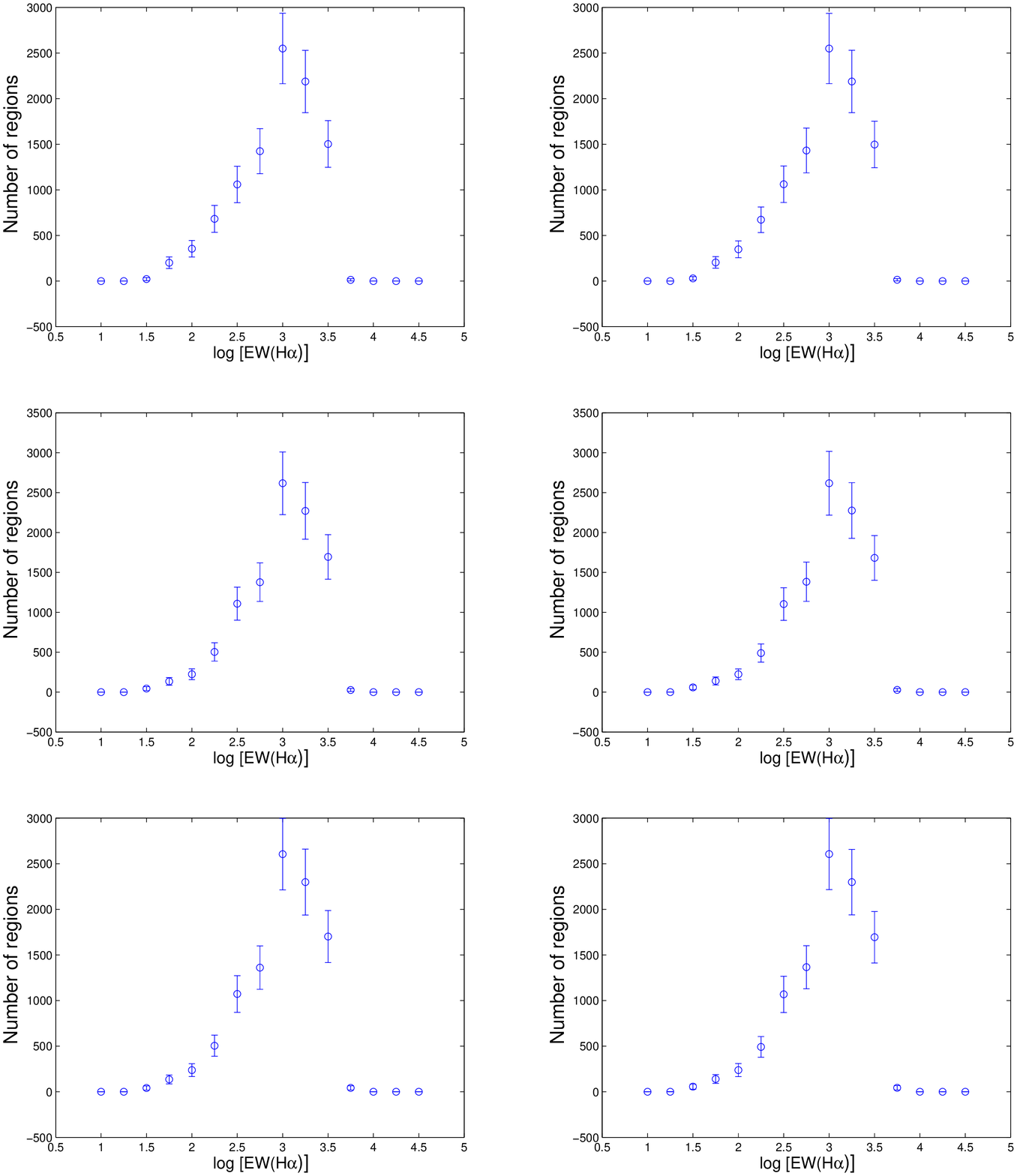,width=176mm}
\caption{Mean of 1000 Monte Carlo simulations for NGC~5457 with photon leakage. Models are: top left, model 1; top right, model 2;
 middle left, model 3; middle right, model 4; bottom left, model 5; bottom right, model 6. Numbers as in Table~\ref{escgd}}
\label{mosegd}
\end{figure}

In Figure~\ref{mosegd} we represent the mean results of the 1000 simulations employing the parameters sumarized in Table~\ref{escgd}.
 The error bars here were calculated as an addition of the statistical error ($3\sigma$) plus the uncertainty in the conversion factor
  between luminosity and star's ionizing mass from Kennicutt (1988) plus the uncertainty in the scaling factor between Leitherer 
  et al.'s (1999) models of ionizing photons and our data.

\begin{table}
\begin{tabular}{c|c|c|c|c|c|c}
\hline
\multicolumn{3}{c|}{Met.} & Escape & 2 TEST & Mean & Number\\
 2Z$_\odot$ & Z$_\odot$ & 0.4Z$_\odot$ & Law  & & & \\ 
\hline
 0\% & 0\% & 100\% & den. & $<$1\% & 3.02$\pm$0.01 & 1\\
 0\% & 0\% & 100\% & f.f. & $<$1\% & 3.01$\pm$0.01 & 2\\
 0\% & 100\% & 0\% & den. & $<$1\% & 2.97$\pm$0.01 & 3\\
 0\% & 100\% & 0\% & f.f. & $<$1\% & 2.97$\pm$0.01 & 4\\
 100\% & 0\% & 0\% & den. & $<$1\% & 2.88$\pm$0.01 & 5\\
 100\% & 0\% & 0\% & f.f. & $<$1\% & 2.88$\pm$0.01 & 6\\ 
\hline
\end{tabular}
\caption{Parameters of the Monte Carlo simulations for NGC~4395 with photon leakage. Columns 1 to 3 are the percentage of 
 regions with a determinate metallicity. The column 1 is the percentage of regions with a metallicity twice  solar. 
 Column 2 is the percentage of regions with a metallicity equal to the solar value. Column 3 is the percentage of regions with a 
 metallicity 0.4 times the solar value. Column 4 is the photon escape law employed from Zurita et al.\ (2002). ''Den'' indicates 
 constant density and ``f.f.'' indicates constant filling factor. Column 5 is the confidence of the two sample test when comparing 
 the model with the data. The higher the percentage, the more similar the simulation to the real data. Column 6 is the mean
  value of the distribution. Column 7 is the identification number of the model. A Salpeter IMF was employed for all the models.}
\label{escff}
\end{table}

In Table \ref{escff} are sumarized the results of the simulations for the flocculent galaxy NGC~4395. For all the possible 
combinations of metallicities, the coincidence between the simulated and the real data is less than 1\%. In accordance with this 
result it is not possible to reproduce the behavior of the H$\alpha$ equivalent width of NGC~4395 by taking  into account
 photon leakage effects alone.

\begin{figure*}
\epsfig{file=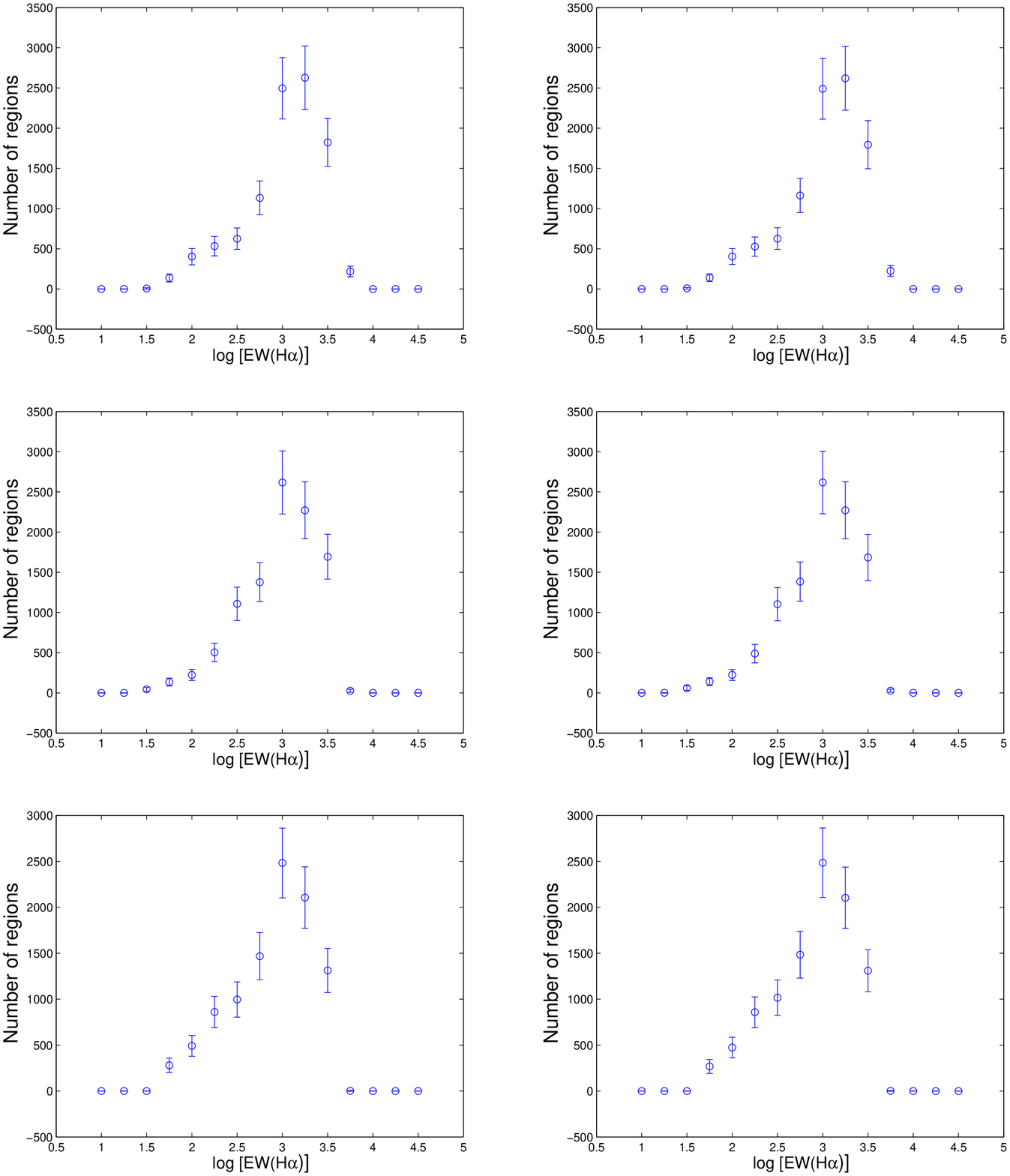,width=176mm}
\caption{Mean of 1000 Monte Carlo simulations for NGC~4395 with photon leakage. Models are: top left, model 1; top right, 
model 2; middle left, model 3; middle right, model 4; bottom left, model 5; bottom right, model 6. Numbers as in Table~\ref{escff}.}
\label{mosefl}
\end{figure*}

In figure~\ref{mosefl} we have represented the mean results of 1000 simulations with the parameters sumamrized in Table~\ref{escff}. 
The error bars were estimated in the same way as in Figure~\ref{mosegd}.

\section{Conclusions}
We have developed a comparative study between the H$\alpha$ equivalent widths of a flocculent galaxy (NGC~4595) and
a grand design galaxy (NGC~5457).

We have found that the H$\alpha$ equivalent width for NGC~5457 is slightly larger than for NGC~4395. In spite of the wide 
dispersion of the data, 
a statistical test based on a two-sample test and a $\chi^2$ test confirm this difference.

In the present paper it is shown that the H$\alpha$ equivalent width difference cannot be attributed only to the age, the metallicity 
or the photon leakage of the H~{\sc ii} regions of the galaxies.

Monte Carlo tests suggest that it is possible that this difference could be cause also by a difference in the IMF, with the flocculent galaxy 
being somewhat deficent in massive stars compared with the grand design galaxy. Furthermore, even if the distribution of the H$\alpha$
 equivalent width for the grand design galaxy could be explained only by employing a Salpeter IMF and a photon leakage law, this is not the
  case for the flocculent galaxy, where only a truncated Salpeter IMF is able to reproduce the results. It is possible that this 
  difference could be generated by a stronger density wave in the grand design galaxy that favors the generation of more massive stars. This not surprising because, as stated in the introduction of this paper, previous studies (ie: McCrady et al. 2003, Smith \& Gallagher 2001, Alonso-Herrero et al. 2001) suggest that there is a possibility of the presence of more massive stars in higher density regions. Moreover, the density wave in the grand design galaxy seems to act in such a way that affects the whole disk, including arm as well as inter-arm regions, indicating that the density wave passage has long-term effects on molecular clouds.

Though these conclusions come through a careful analysis of the two galaxies, the impossibility of reproducing with sufficient accuracy the equivalent width distribution for the flocculent galaxy and the 
multiple factors occurring simultaneously (different SFRs and photon leakage) makes it neccesary to carry out more observations 
in different grand design and flocculent galaxies in order to confirm this behavior.

\acknowledgements
We thank the anonymous referee for most helpful comments and suggestions. We also thank Dr. Jorge Iglesias and Dr. H\'ector Casta\~neda for their suggestions to improve this paper.
B.C. wishes to thank the staff of Wakayama University for their inmense help during the first stages of this paper.
Part of this work was supported by the Spanish Plan Nacional de Astronom\'{\i}a y Astrof\'{\i}sica under grant AYA2002-01379.


\begin{thebibliography}{}
\bibitem[2001]{669} Alonso-Herrero A., Engelbracht C.W., Rieke M.J., Rieke G.H., \& Quillen A.C. 2001, ApJ 546, 952
\bibitem[1992]{671} Belley, J., \& Roy, J. 1992, ApJS, 78, 61
\bibitem[1997]{672} Bresolin, F., \& Kennicutt, R.C., Jr. 1997, AJ, 113, 975
\bibitem[2002]{bj} Cedr\'es, B., \& Cepa, J. 2002, A\&A, 391, 809
\bibitem[2004]{bmj} Cedr\'es, B., Urbaneja, M.A., \& Cepa, J. 2004, A\&A, 422, 511
\bibitem[1989]{675} Cepa, J., \& Beckman, J.E. 1989, Ap\&SS, 156, 289
\bibitem[1990]{676} Cepa, J., \& Beckman, J.E. 1990a, ApJ, 349 497
\bibitem[1990]{677} Cepa, J., \& Beckman, J.E. 1990b, Ap\&SS, 170, 209
\bibitem[1986]{679} Coppetti, M.V.F., Pastoriza, M.G., \& Dottori, H.A. 1986, A\&A, 156, 111
\bibitem[1991]{680} de Vaucouleurs, G., de Vaucouleurs, A., Corwin, H. G., et al. 1991, Third Reference Catalogue of Bright Galaxies 
(New York: Springer)  
\bibitem[1985]{682} Efremov, Yu.N. 1985, SvA, L11, 169
\bibitem[1992]{683} Elmegreen, B.G. 1992 {\it Star Formation in Stellar Systems}, Cambridge Univ. Press
\bibitem[2004]{684} Elmegreen, B.G. 2004, MNRAS 354, 367
\bibitem[1985]{685} Elmegreen, B.G., \& Elmegreen, D.M. 1984, ApJSS, 54, 127
\bibitem[1986]{686} Elmegreen, B.G., \& Elmegreen, D.M. 1986, ApJ, 311, 554
\bibitem[1987]{687} Elmegreen, D.M., \& Elmegreen, B.G. 1987, ApJ, 314, 3
\bibitem[2003]{689} F\"orster Schreiber N.M., Genzel R., Lutz D, \&. Sternberg A. 2003, ApJ 599, 193
\bibitem[2002]{690} Gavazzi, G., Boselli, A., Pedotti, P., Gallazzi, A., \& Carrasco, L. 2002, A\&A, 396, 449
\bibitem[2005]{693} Gonz\'alez P\'erez, J.M.,\& Cepa J. 2005, in preparation.
\bibitem[1990]{695} Hodge, P., Jaderlund, E., \& Meakes, M. 1990, PASP, 102, 1263
\bibitem[2003]{696} Karachentsev, I.D., Sharina, M.E., Dolphin, A.E., Grebel, E.K., Geisler, D., Guhathakurta, P., Hodge, P.W., 
Karachentseva, V.E., Sarajedini, A., \& Seitzer, P. 2003, A\&A, 398, 467
\bibitem[1996]{698} Kelson, D.D., Ilingworth, G.D., Freedman, W.F., Graham, J.A., Hill, R., Madore, B.F. \& 12 coautors. 1996, ApJ, 463, 26
\bibitem[1983]{699} Kennicutt, R.C., Jr. 1983, ApJ, 272, 54
\bibitem[1988]{700} Kennicutt, R.C., Jr. 1988, ApJ, 334, 144
\bibitem[2003]{701} Kennicutt, R.C., Jr., Bresolin, F., \& Garnett, D.R. 2003, ApJ, 591, 801
\bibitem[1989]{702} Kennicutt, R.C., Jr., Keel, W.C., \& Hodge, P.W. 1989, ApJ, 337, 761
\bibitem[1994]{703} Kennicutt, R.C., Jr., Tamblyn, P., \& Congdon, C.W. 1994, ApJ, 435, 22
\bibitem[1992]{704} Knapen, J.H., Beckman, J.E., Cepa, J., van der Hulst, T., \& Rand, R.J. 1992, ApJ, 385, 37
\bibitem[1999]{706} Leitherer, C., Schaerer, D., Goldader, J.D., Gonz\'alez Delgado, R.M., Robert, C., Foo Kune, D., De Mello, D.F., 
Devost, D., \& Heckman, T.M. 1999, ApJS, 123, 3
\bibitem[1964]{708} Lin C.C.,\& Shu F.H. 1964, ApJ 140, 646
\bibitem[1999]{709} Martin, P., \& Friedli, D. 1999, A\&A, 346, 769
\bibitem[1986]{710} McCall, M.L., \& Schmidt, 1986, AJ, 311, 548
\bibitem[2003]{711} McCrady N., Gilbert A.M., \& Graham J.R. 2003, ApJ 596, 240
\bibitem[1989]{712} Osterbrock, D.E. 1989, Astrophysics of Gaseous Nebulae and Active Galactic Nuclei (University Science Books)
\bibitem[1996]{715} Roy, J., Belley, J., Dutil, Y., \& Martin, P. 1996, ApJ, 460, 284
\bibitem[1982]{716} Seiden, P.E., \& Gerola, H. 1982, Fund. Cosm. Phys., 7, 241
\bibitem[2004]{717} Shadmehri M. 2004, MNRAS 354, 375
\bibitem[1989]{718} Sitnik, T.G. 1989, SvA, L15, 338
\bibitem[1991]{719} Sitnik, T.G. 1991, SvA, L17, 61
\bibitem[2001]{720} Smith L.J., \& Gallagher J.S., 2001, MNRAS 326, 1027
\bibitem[1998]{721} Sternberg A. 1998, ApJ 506, 721
\bibitem[1990]{722} Tacconi, L.J., \& Young, J.S. 1990, ApJ, 352, 595
\bibitem[1998]{723} Van Zee, L., Salzer, J.J., Haynes, M.P., O'Donoghue, A.A., \& Balonek, T.J. 1998, AJ, 116, 2805
%\bibitem[1991]{724} Wilson, C.D., \& Scoville, N. 1991, AJ. 101, 1293
\bibitem[2002]{725} Zurita, A., Beckman, J.E., Rozas, M., \& Ryder, S. 2002, A\&A, 386, 801

\end{thebibliography}
\end{document}